\documentclass[fleqn,12pt,twoside]{article}
\usepackage[headings]{espcrc1}

\readRCS
$Id: espcrc1.tex,v 1.2 2004/02/24 11:22:11 spepping Exp $
\usepackage{graphicx}
\usepackage[figuresright]{rotating}

\newcommand{\AmS}{{\protect\the\textfont2
  A\kern-.1667em\lower.5ex\hbox{M}\kern-.125emS}}

\usepackage{amssymb}

\title{Close-coupling calculations of rotational energy transfer in $p$-H$_2$+HD} 

\author{Renat A. Sultanov\address[BCRL]{Business Computing Research Laboratory, St. Cloud State University,
31 Centennial Hall, 720 Fourth Avenue South, St Cloud, MN 56301-4498}
\thanks{rasultanov@stcloudstate.edu; r.sultanov2@yahoo.com},
Dennis Guster\addressmark\thanks{dcguster@stcloudstate.edu}.}

\runtitle{Rotational energy transfer in $p$-H$_2$+HD}
\runauthor{R.A. Sultanov and D. Guster}

\begin{document}

\maketitle

\begin{abstract}
We report quantum-mechanical close-coupling calculations for rotational state resolved cross sections 
for $p$-H$_2$+HD collisions.
The low temperature limit of $p$-H$_2$+HD is investigated, which is of significant
astrophysical interest in regard to the
cooling of primordial gas and the interstellar media. Sharp resonances 
have been reproduced in the cross sections of some transition states at very low kinetic energies,
$E_{kin}\sim 10^{-5}$ eV.
\end{abstract}


\section{Introduction}

Energy transfer collisions between H$_2$ and HD molecules, where H is hydrogen and D is deuterium,
is of fundamental importance
for the astrochemistry of the early Universe and the interstellar medium
\cite{dalgar72,shih75,schaefer90,puy93,galli98,stancil98,flower99,flower99a,flower00,flower00a,flower01,galli02,lepp02}.
HD represents the second most abundant
primordial molecule after H$_2$ and plays a significant role in the cooling of the primordial gas.
The abundance of deuterium is low relative to hydrogen (D/H$\approx10^{-5}$), however
the HD/H$_2$ ratio can be enhanced by an approximate factor of $10^2$ due to chemical fractionation
\cite{puy93,stancil98}.

It has been shown, that
in the framework of the standard cosmological model the radiation temperature is higher than the matter
temperature and molecules become a heating source for the gas.
At higher temperatures H$_2$ molecules dominate the 
heating, however HD molecules dominate the kinetic process at lower temperatures.
In addition, the HD molecule is especially important due to its permanent dipole moment and
lower rotational constant, which makes the molecule to be an efficient coolant at low temperatures:
$T \lesssim 100$K \cite{lepp02}. 
When H$_2$ molecules are inefficient HD become important in cooling the primordial gas and  the
interstellar medium. This ability of HD makes these molecules to be very attractive.

Therefore knowledge of the ro-vibrational excitation and de-excitation thermal rate constants in 
the molecular H$_2$+HD low energy collisions 
is of fundamental importance in understanding and modeling the energy balance within primordial gas. 
However, to accurately model the thermal balance and kinetics of such systems one needs accurate state-to-state 
cross-sections and rate constants $k_{vjv'j'}(T)$.

Experimental measurement of quantum state resolved cross sections and rate coefficients is a very difficult 
technical problem. On the other hand, accurate theoretical data require precise potential energy surfaces
and reliable dynamical treatment of the collision processes.
The first attempt to construct a realistic full-dimensional ab initio PES for the H$_2$-H$_2$
system was done in works \cite{schwenke88,schwenke90}, 
and the potential was widely used in a variety of methods and computation techniques.
Currently the deuterium chemistry of the eraly Universe has been extensively studied by many researchers.
Here we would like to mention works \cite{flower99,flower99a} and \cite{schaefer90},
where the authors carried out quantum-mechanical state-resolved calculations for low and very low temperatures
down to 10 K \cite{schaefer90}, however using of the old and modifyed H$_2$-H$_2$ surface \cite{schwenke88}.

Nontheless the importance of the H$_2$+HD 
in the astrophysical problems makes it vital to carry out
new calculations for the system with recently published PES \cite{booth02}. This is done here in. 
Moreover, with the current calculations we extend our recent experience for H$_2$+H$_2$ at low and very
low kinetic energies \cite{renat06a,renat06b}.

As was mentioned above
a new extensive study of the hydrogen-hydrogen surface has been recently reported by Boothroyd et al.
\cite{booth02}, where the potential energies have been represented at 48180 geometries respectively
with a large basis set at the multireference configuration interaction level. 
In this work we provide the first calculations describing collisions of rotationally excited H$_2$ 
and HD molecules  using the BMKP PES

\begin{equation}
p\mbox{-H}_2(j_2) +\mbox{HD}(j_1) \rightarrow p\mbox{-H}_2(j'_2) + \mbox{HD}(j'_1).
\label{eq:h2hd}
\end{equation}

The scattering cross sections 
are calculated using a non reactive quantum-mechanical close-coupling
approach. In the next section we will briefly outline the method and present results for the cross sections
for rotational de-excitation of HD in low energy collisions with $p$-H$_2$. We compare our results with
some previous investigations. Conclusions are presented in Section 3.

\section{Method and Results}

In this section we provide a brief outline of the quantum-mechanical close-coupling approach used 
in our calculations. 
The basis for this methodology was developed in work \cite{green75}.
The HD and H$_2$ molecules are treated as linear rigid  rotors. 
For the considered range of kinetic energies of astrophysical interest 
and for practical astrophysical estimations
the rotor model is considered to be adequate for $p$-H$_2$+HD collisions \cite{flower99a}.

The 4-atomic H$_2$-HD system is shown in Fig. 1. It can be described by six independent variables:
$R_1$ and $R_2$ are interatomic distances in each hydrogen
molecule,  $\Theta_{1}$ and $\Theta_{2}$ are polar angles,  $\Phi_2$ is torsional angle and 
$R_3$ is intermolecule distance.
Let us introduce $M_{12} = (m_1+m_2)(m_3+m_4)/(m_1+m_2+m_3+m_4)$ and 
$\mu_{1(2)}=m_{1(3)}m_{2(4)}/(m_{1(3)}+m_{2(4)})$, 
where the first one is a reduced mass of the pair of two-atomic molecules 
$12$ and $34$ and  the second ones are reduced masses in hydrogen molecules.
As we mentioned, the hydrogen molecules are treated as 
linear rigid rotors, that is distances $R_1=0.7631$\ a.u. in H$_2$ and $R_2=0.7668$\ a.u. in HD
are fixed in this model.
We provide a numerical solution for the Schr\"odinger equation for a $12+34$ collision in the center of the
mass frame.

The cross sections for rotational excitation and relaxation phenomena can be obtained directly 
from the $S$-matrix.
In particular, the cross sections for excitation from $j_1j_2\rightarrow j'_1j'_2$ summed over 
the final $m'_1m'_2$
and averaged over the initial $m_1m_2$ corresponding projections of the H$_2$ molecules angular 
momenta $j_1$ and $j_2$ are given by
\begin{eqnarray}                               
\sigma(j'_1,j'_2;j_1j_2,\epsilon)=\frac{\pi}{(2j_1+1)(2j_2+1)k_{\alpha\alpha'}}  
\sum_{Jj_{12}j'_{12}LL'}(2J+1)|\delta_{\alpha\alpha'}-
S^J_{\alpha \alpha'}(E)|^2.
\label{eq:cross}
\end{eqnarray}
The kinetic energy is 
%
$\epsilon=E-B_1j_1(j_1+1)-B_2j_2(j_2+1)$.
%
Here $E$ is the total energy in the system, $B_{1}=60.8\hspace{2mm}\mbox{cm}^{-1}$ and
$B_{2}=44.7\hspace{2mm}\mbox{cm}^{-1}$
are the rotation constants of the colliding HD and H$_2$ molecules respectively,
$J$ is total angular momenta of the 4-atomic system, 
$\alpha \equiv (j_1j_2j_{12}L)$, 
where $j_1+j_2=j_{12}$ and $j_{12}+L=J$,
$k_{\alpha \alpha'}=2M_{12}(E+E_{\alpha}-E_{\alpha'})^{1/2}$ is the channel wavenumber and
$E_{\alpha(\alpha')}$ are rotational channel energies.

We apply the hybrid modified log-derivative-Airy propagator in the general purpose scattering program MOLSCAT 
\cite{hutson94} to solve a set of coupled second order differential equations
for the unknown radial functions $U^{JM}_{\alpha}(R)$
\begin{eqnarray}
\left(\frac{d^2}{dR^2}-\frac{L(L+1)}{R^2}+k_{\alpha}^2\right)U_{\alpha}^{JM}(R)=2M_{12}
\sum_{\alpha'} \int <\phi^{JM}_{\alpha}(\hat r_1,\hat r_2,\vec R)\nonumber \\
|V(\vec r_1,\vec r_2,\vec R)| 
\phi^{JM}_{\alpha'}(\hat r_1,\hat r_2,\vec R)>U_{\alpha'}^{JM}(R) d\hat r_1 d\hat r_2 d\hat R,
\label{eq:cpld}
\end{eqnarray}
%
%
We have tested other propagator schemes included in the MOLSCAT code. It was found, that
other propagators can also produce quite stable results.

The log-derivative matrix is propagated to large $R$-intermolecular distances, since all experimentally observable
quantum information about the collision is contained in the asymptotic behaviour of functions 
$U^{JM}_{\alpha}(R\rightarrow\infty)$. The numerical results are matched to the known asymptotic solution to 
derive the physical scattering $S$-matrix
%
%
\begin{equation}
U_{\alpha}^J
\mathop{\mbox{\large$\sim$}}\limits_{R \rightarrow + \infty}
\delta_{\alpha \alpha'}
e^{-i(k_{\alpha \alpha}R-(l\pi/2))} 
- \left(\frac{k_{\alpha \alpha}}{k_{\alpha \alpha'}}\right)^{1/2}S^J_{\alpha \alpha'}
e^{-i(k_{\alpha \alpha'}R-(l'\pi/2))},
\end{equation}
where $k_{\alpha \alpha'}=2M_{12}(E+E_{\alpha}-E_{\alpha'})^{1/2}$ is the channel wavenumber, 
$E_{\alpha(\alpha')}$
are rotational channel energies and $E$ is the total energy in the $1234$ system.
The method was used for each partial wave until a converged cross section was obtained. 
It was verified that the results are converged with respect to the number of partial waves as well as
the matching radius, $R_{max}$, for all channels included in our calculations.

The new BMKP PES \cite{booth02}, which is used in these calculations,
is a global six-dimensional potential energy surface for two hydrogen molecules.
It was especially constructed to represent the whole interaction region of the chemical reaction dynamics 
of the  four-atomic system and to provide an accurate as possible van der Waals well.
In the six-dimensional conformation space of the four atomic system the conical intersection forms a 
complicated three-dimensional hypersurface. 

Because the BMKP PES \cite{booth02}
uses cartesian coordinates to compute the distances between
four atoms, we needed to devise a fortran program, which converts spherical coordinates used in
the close coupling method \cite{hutson94} to the corresponding cartesian coordinates and computes 
the distances between the four atoms followed by calculations of interatomic interaction forces.

The four atomic system is shown in Fig.\ 1. Let us introduce
the Jacobi coordinates $\{\vec R_1, \vec R_2, \vec R_3\}$ and the radius-vectors of all four atoms in
the space-fixed coordinate system $OXYZ$: $\{\vec r_1, \vec r_2, \vec r_3, \vec r_4\}$.
We apply the following procedure:
$\vec R_3$ is directed along $OZ$ axis,
the center of mass of the HD molecule is brought into coincidence with the center of $OXYZ$, and the top of the
$\vec R_3$ is directed to center of mass of H$_2$, as shown in Fig.\ 1.
Now it is apparent, that $\vec R_3=\{R_3, \Theta_3=0, \Phi_3=0\}$,
$\vec R_1=\vec r_1 - \vec r_2$, $\vec R_2=\vec r_4 - \vec r_3$, and
$\vec r_1=\alpha \vec R_1$ and $\vec r_2=(1 - \alpha) \vec R_1$, where $\alpha=m_2/(m_1+m_2)$. 
Next, without the loss of generality, we can adopt the $OXYZ$ system in such a way, that
the HD interatomic vector $\vec R_1$ lies on the $XOZ$ plane. Then the angle variables of 
$\vec R_1$ and $\vec R_2$ are: $\hat R_1=\{\Theta_1,\Phi_1=\pi\}$
and $\hat R_2=\{\Theta_2,\Phi_2\}$ respectively.

Now one can see, that the cartesian coordinates of the atoms of the HD molecule are:
$\vec r_1 = \{x_1=\alpha R_1\sin\Theta_1, y_1=0, z_1=\alpha R_1 \cos\Theta_1\}$,
$\vec r_2 = \{x_2=(1-\alpha) R_1 \sin\Theta_1, y_2=0, z_2=-(1-\alpha) R_1 \cos\Theta_1\}$, and
in turn for the H$_2$ molecule we have:
$\vec r_3 = \{x_3=-(1-\beta) R_2 \sin\Theta_2 \cos\Phi, y_3=-(1-\beta)R_2\sin\Theta_2\sin\Phi,
z_3=R_3-(1-\beta)R_2\cos\Theta_2\}$ and
$\vec r_4 = \{x_4=\beta R_2 \sin\Theta_2 \cos\Phi, y_4=\beta R_2 \sin\Theta_2\sin\Phi,
z_4=R_3+\beta R_2\cos\Theta_2\}$, because 
$\vec r_3 = \vec R_3 - (1-\beta)\vec R_2$ and $\vec r_4 = \vec R_3 + \beta \vec R_2$, 
where $\beta = m_4/(m_3+m_4)$.
In such a manner the cartesian and the Jacobi coordinates are represented together for the four-atomic system
H$_2$+HD.

Before our production calculations we carried out a
large number of test calculations to insure the convergence of the results with respect to all 
parameters that enter into the propagation of the Schr\"odinger equation. The same calculations were
also done in our previous works \cite{renat06a,renat06b} for the $o$-/$p$-H$_2$+H$_2$ collisions, 
which involved     
the intermolecular distance $R$, the total angular momentum $J$ of the four atomic system, 
the number of rotational levels to be included in the close coupling expansion $N_{lvl}$, and others
(see the MOLSCAT manual \cite{hutson94}).

We reached convergence for the integral cross sections, $\sigma(j'_1,j'_2;j_1j_2,\epsilon)$,
in all considered collisions.
For example, for $R$ we used from $R_{min}=1$ \r{A} to $R_{max}=22$ \r{A}, we also applied a few 
different propagators included in the MOLSCAT program. We obtained convergent results in all cases.

Below we present our calculations for the state-resolved cross sections 
$\sigma(j'_{1}j'_{2};j_{1}j_{2})(\epsilon)$
in the collision (1). Fig.\ 2 shows our data for  the $(j_1=1,j_2=0)\rightarrow(j'_1=j'_2=0)$ 
and $(j_1=2,j_2=0)\rightarrow(j'_1=1,j'_2=0)$ 
quantum transitions together with the results of Schaefer's work \cite{schaefer90},
which applied a modified version of Schwenke's potential 
\cite{schwenke88} and different dynamical quantum-mechanical
approach. As can be seen for these transition states the agreement between the two different 
calculations is excellent.
We reproduced sharp resonances in the low velocity region, which are very important in cooling of the
astrophysical media. In addition to Fig.\ 2 we separately show the cross section for only 
$(j_1=1,j_2=0)\rightarrow(j'_1=j'_2=0)$ transition states in Fig.\ 3.
This graph can be directly compared with the
corresponding cross section from the Schaefer work \cite{schaefer90}, please refer to
figure 6 of that paper \cite{schaefer90}.
One can see, that even small peculiarities at around 250 m/s and 450 m/s
of the cross section behaviour are reproduced in details.

In the Fig. 4 we present the energy dependence of the cross sections for the
$(j_1=2,j_2=0)\rightarrow(j'_1=j'_2=0)$ 
and
$(j_1=1,j_2=2)\rightarrow(j'_1=0,j'_2=2)$ 
quantum-mechanical transition states.
For the last transition we obtain very good agreement with the corresponding results of
work \cite{schaefer90}. However, in the upper graph of Fig.\ 4  one can
see rather large differences in the low velocity region. Our calculations revealed, that
the disagreement ranges up to about 100 \%.

Finally, significant differences are found in the cross sections
from the higher values of the transition states.  Fig.\ 5 presents our results again
together with Schaefer's data from \cite{schaefer90} for
$(j_1=1,j_2=2)\rightarrow(j'_1=2,j'_2=0)$,
$(j_1=1,j_2=2)\rightarrow(j'_1=1,j'_2=0)$
and
$(j_1=1,j_2=2)\rightarrow(j'_1=j'_2=0)$.
%
%
In this case the disagreement ranges up to one order of magnitude.

Also, as can be seen from the graphs we calculated few new resonances for each of these transitions
at a velocity of about 1200 m/s. The biggest value of these resonances is in the cross section
$(j_1=1,j_2=2)\rightarrow(j'_1=2,j'_2=0)$.
We show this separately in more detail in Fig.\ 6. The value of the resonance is relatively large and it
also might be applicable in important astrophysical processes, such as collisional cooling.

Through this analysis we can now conclude, that the new global BMKP PES can reproduce general behaviour 
of all considered cross sections in the $p$-H$_2$+HD collision. For the lower quantum states we obtained
sufficient agreement with previous calculations.
However, for transition states from higher values, for example, $(j_1=2,j_2=0)$ the BMKP PES provides 
rather low cross sections relatively to the corresponding results of work \cite{schaefer90}.

\section{Conclusion}

In this letter the state-to-state close-coupling quantum-mechanical calculations for rotational
excitation and deexcitation cross sections of the $p$-H$_2$+HD collision are presented.
The linear rigid rotor model for the H$_2$ and HD molecules is applied.
The global and newest BMKP surface for the H$_2$-H$_2$ system has been appropriately adopted for the current
$p$-H$_2$+HD collisions. A test of convergence and the results for cross sections with the BMKP PES
are obtained for a wide range of kinetic velocities including very low values down to 10 m/s.

Our results revealed, that for low quantum transition states the BMKP surface provides cross sections very
close to those obtained in previous works \cite{schaefer90,flower99}, where the authors
adopted Schwenke's old H$_2$-H$_2$ PES \cite{schwenke88}.
However, for some higher quantum states
we found significant disagreements with previous results \cite{schaefer90} (see Fig.\ 5). Additionally, 
in our calculations some
new resonances are found in the 1300$\pm 100$ m/s region for transition states 
from $j_2=2$ and $j_1=1$. It was found, that for the specific transition 
$(j_2=2,j_1=1)\rightarrow (j_2=2$, $j_1=0)$ the value of the resonance is relatively large in the cross section,
and it may even stronger influence, for example, the cooling processes in primodial gas and interstellar media.
Further detailed calculations for higher quantum transition state cross sections and corresponding
thermal rate constants $k_{j_1j_2\rightarrow j'_1j'_2}(T)$ for $o$-/$p$-H$_2$+HD collisions at low and 
very low kinetic energies are in progress in our group.

In conclusion, we would like to point out here, that
the ultralow energy sector 
$(T \lesssim 1\ \mu$K)
is of crucial importance now
in connection with the recently achieved
molecular Bose-Einstein condensates \cite{greiner03,jochim03,zwier03,demille02,hope01,heinzen00}.
Therefore, it should also be interesting and useful to apply the current time-independent, quantum-mechanical
approach to investigate isotope effects in molecular hydrogen collisions at such ultralow energies.
Further, the methodology could be used to carry out new calculations for the important
ultralow collisions considered in the works \cite{avd06,avd02,avd01}, as well as
check the conclusions of the recent work \cite{petrov05} involving scattering properties of weakly 
bound dimers of fermionic atoms.



\clearpage
\begin{figure}
\begin{picture}(250,250)(-10,0)
%
\put(175,17){\circle*{23}}
\put(153,0){\footnotesize{(2)}}

\put(152,17){\bf D}
\put(267,-6){\circle*{12}}

\put(257,-27){\footnotesize{(1)}}
\put(275,-23){\bf H}

\thicklines
\put(175,15){\vector(4,-1){87}}   
\thinlines

\put(230,-15){$\vec R_1$}   





\thicklines
\put(212.5,7){\vector(0,1){167}}     
\thinlines

\put(193,145){$\vec R_3$}       

\put(212.5,7){\line(0,1){215}}   

\put(212.5,8){\line(1,0){110}}   

\put(212.5,7){\line(-2,-1){80}}   

\put(212.7,7){\vector(1,4){45.7}}  
\put(212.5,7){\vector(-1,3){45.7}} 


\put(212.5,8){\vector(4,-1){50}}     
\put(212.5,8){\vector(-4,1){31}}     

\put(190,17){$\vec r_2$}
\put(276,-4){$\vec r_1$}

\put(170,100){$\vec r_3$}
\put(250,140){$\vec r_4$}

\put(212.5,7){\circle*{3}}      


\put(200,210){$Z$}
\put(312,12){$Y$}
\put(132,-22){$X$}

\put(206,-7){$O$}             
\put(218,12){$\Theta_1$}
\put(215,185){$\Theta_2$}

\qbezier(205,94)(245,108)(205,105)

\put(205,105){\vector(-1,0){4}}    
\put(195,111){$\Phi_2$}                                  

\put(165,150){\circle*{12}}
\put(146,131){\footnotesize{(3)}}
\put(141,150){\bf H}

\put(260,195){\circle*{12}}
\put(269,176){\footnotesize{(4)}}
\put(270,195){\bf H}

\thicklines
\put(165,150){\vector(2,1){89.2}}            
\thinlines
\put(233,199){$\vec R_2$}
\end{picture}
\vspace{12mm}
\caption{Four-atomic coordinates for the $p$-H$_2(j_2)+$HD$(j_1)$ collision used in this work.
$R_{1}=0.7631$ a.u. and $R_{2}=0.7668$ a.u. are fixed interatomic distances in each hydrogen
molecule HD and H$_2$ respectively,
$\Theta_{1}$ and $\Theta_{2}$ are polar angles of vectors $\vec R_1$ and $\vec R_2$ respectively,
$\Phi_2$ is torsional angle and $R_3$ is the intermolecular vector, which connects
the center of masses of the molecules. Vectors $\vec r_1, \vec r_2, \vec r_3$ and $\vec r_4$ represent the cartesian
coordinates of the four atoms in the space-fixed $OXYZ$ coordinate system.}
\label{fig:fig11}
\end{figure}
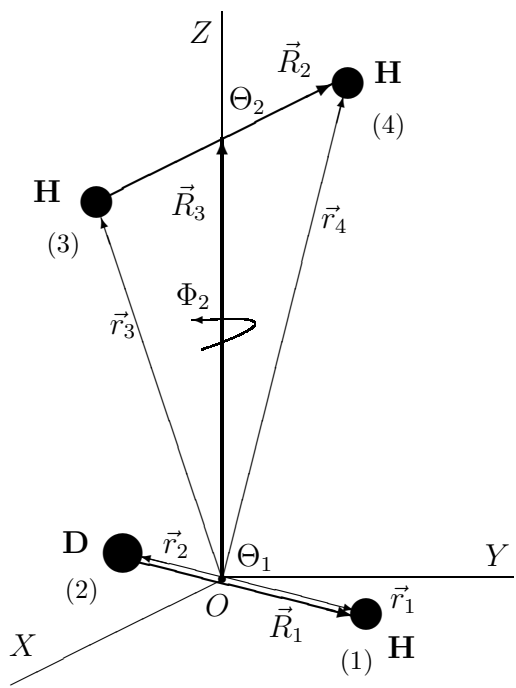


\begin{figure}
\begin{center}
\includegraphics*[scale=1.0,width=27pc,height=17pc]{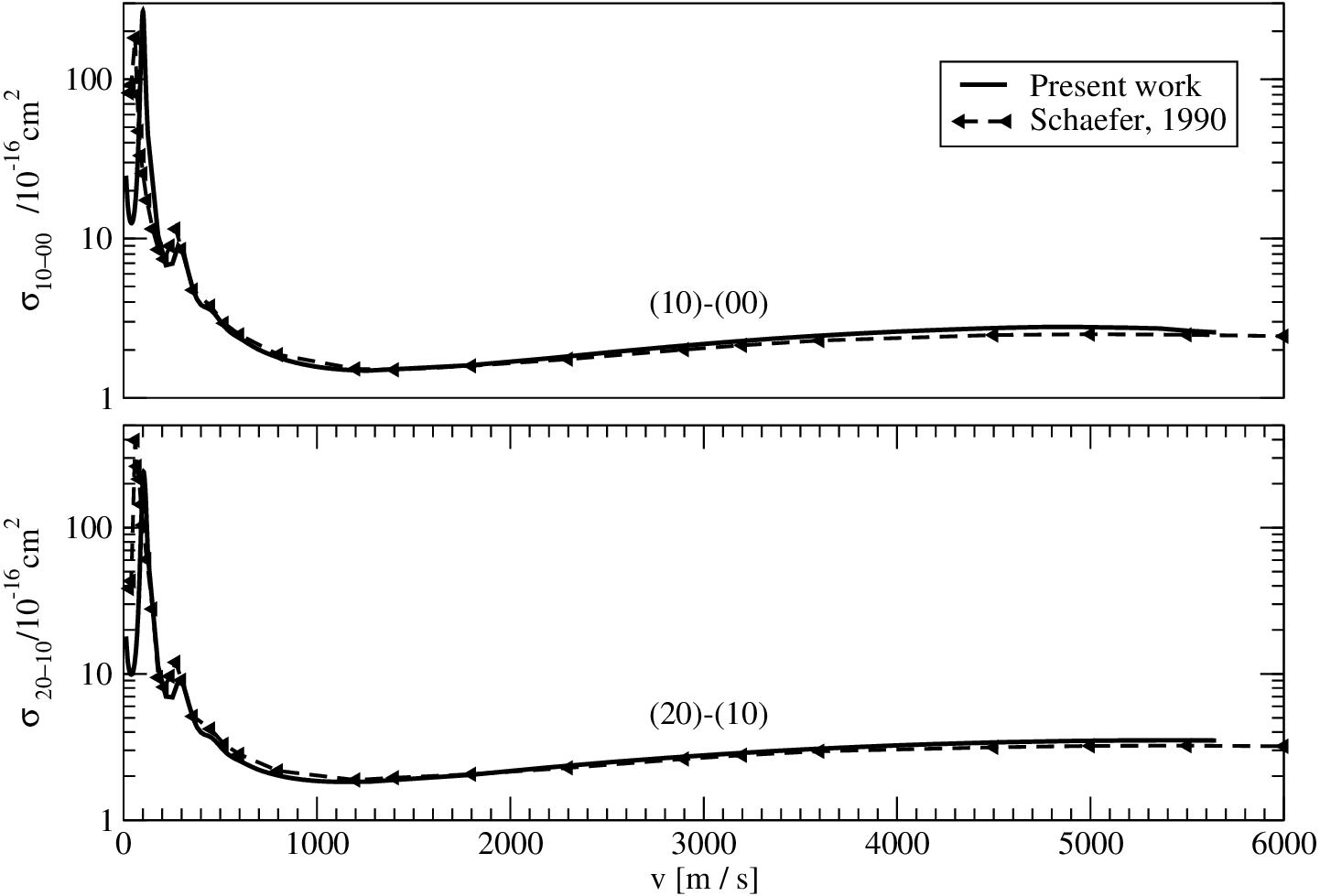}
\end{center}
\caption{
Rotational state resolved integral cross sections for
$p$-$\mbox{H}_2(j_2) +\mbox{HD}(j_1) \rightarrow \mbox{H}_2(j'_2) + \mbox{HD}(j'_1)$.
Initial states of HD and H$_2$ molecules are $j_1=1$ and $j_2=0$ respectively and 
corresponding final states are $j'_1=j'_2=0$. In the bottom plot:
$j_1=2$, $j_2=0$ and $j'_1=1$, $j'_2=0$.
Calculations are done with the BMKP PES (bold lines), triangles left are corresponding 
results from work \cite{schaefer90}.
}
\label{fig:fig1}
\end{figure}       

\begin{figure}
\begin{center}
\includegraphics*[scale=1.0,width=27pc,height=17pc]{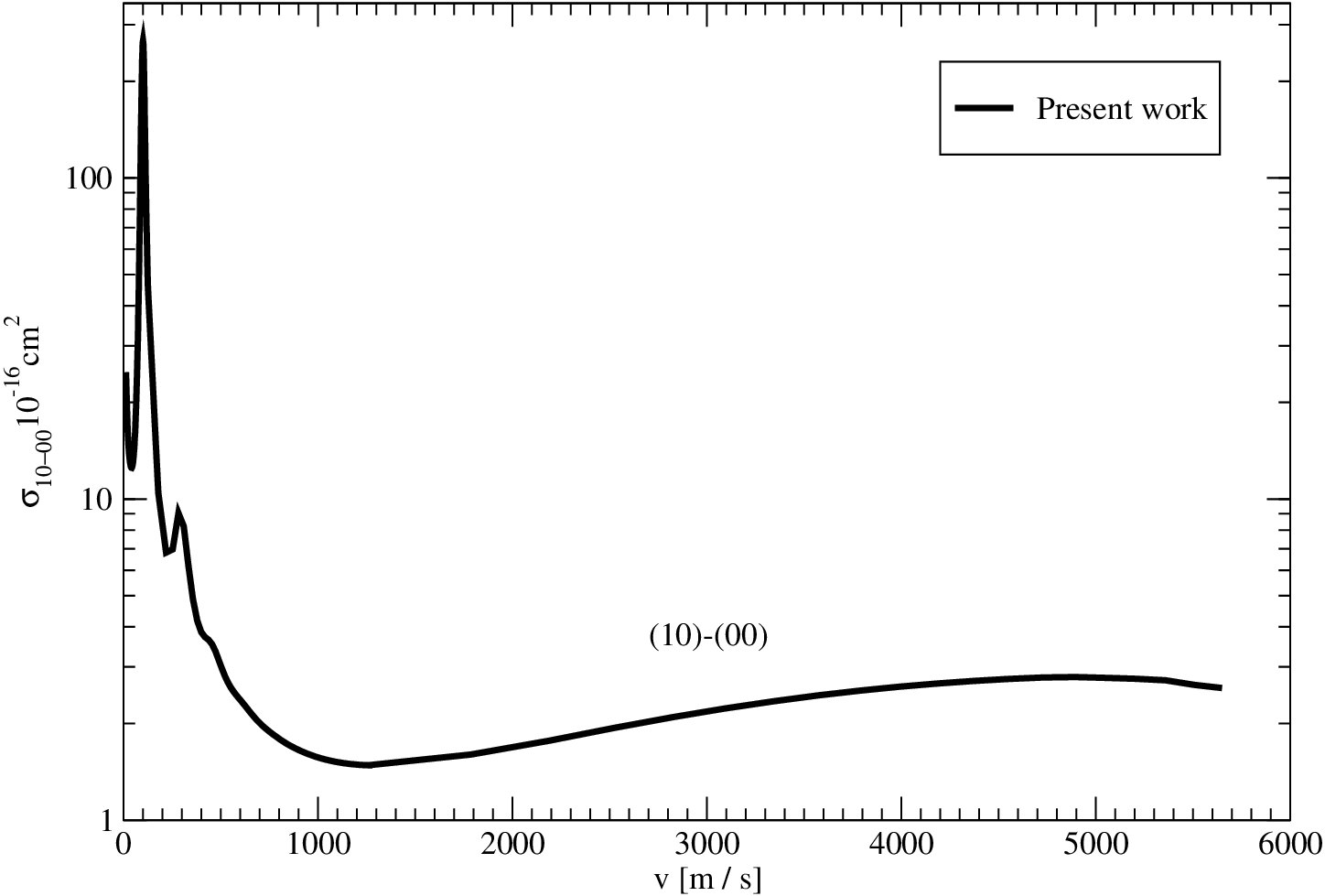}
\end{center}
\caption{Details of the  $(j_1=1,j_2=0)\rightarrow (j'_1=0,j'_2=0)$
rotational transition state cross section in
$p\mbox{-H}_2(j_2=0) +\mbox{HD}(j_1=1) \rightarrow \mbox{H}_2(j'_2=0) + \mbox{HD}(j'_1=0)$
to compare with those corresponding cross sections from work \cite{schaefer90} figure 6.}
\label{fig:fig2}
\end{figure}

\begin{figure}
\begin{center}
\includegraphics*[scale=1.0,width=27pc,height=17pc]{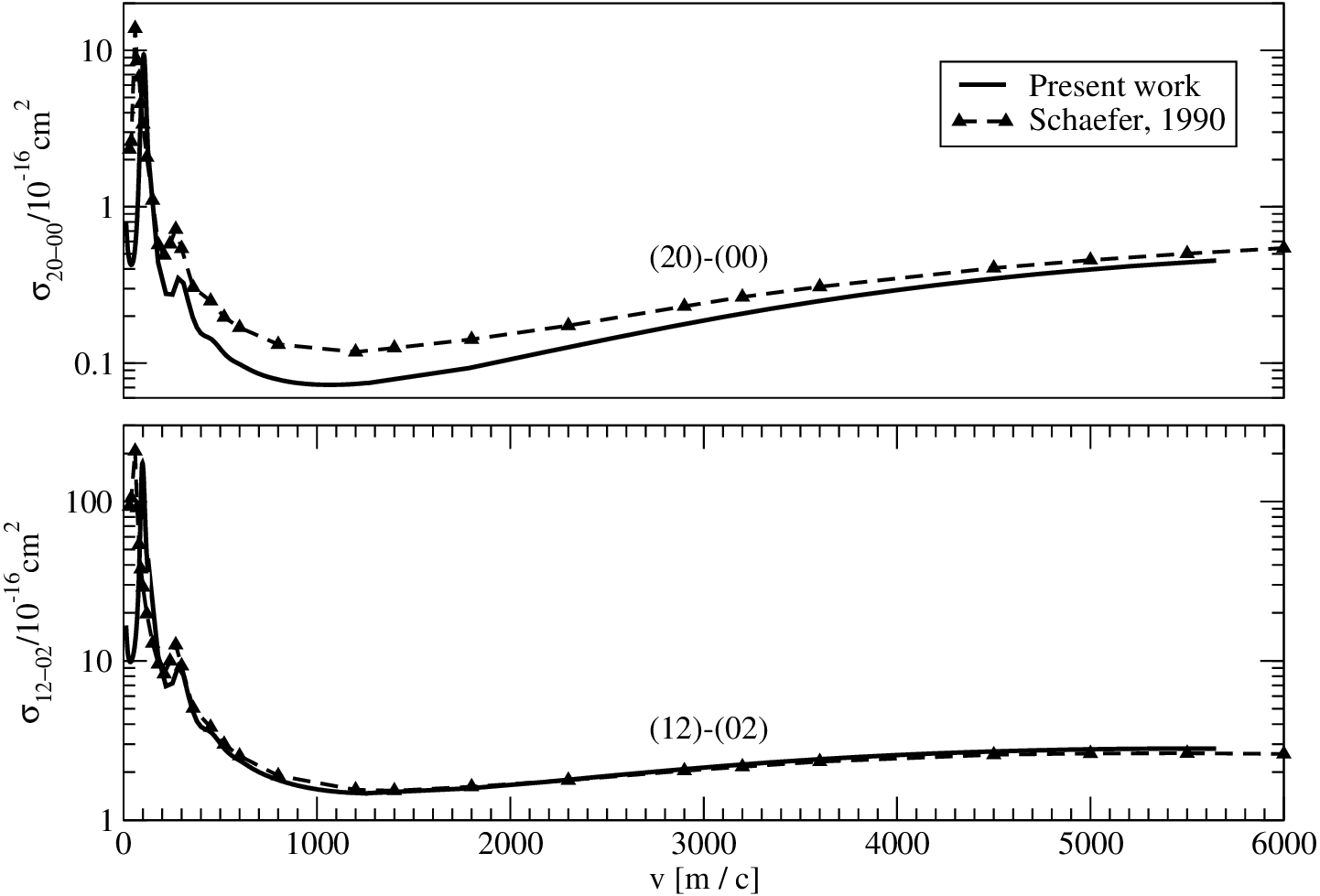}
\end{center}
\caption{Rotational state resolved integral cross sections for
$p$-$\mbox{H}_2(j_2) +\mbox{HD}(j_1) \rightarrow \mbox{H}_2(j'_2) + \mbox{HD}(j'_1)$.
Initial states of HD and H$_2$ molecules are $j_1=2$, $j_2=0$ respectively and 
corresponding final states are $j'_1=j'_2=0$. In the bottom plot:
$j_1=1$, $j_2=2$ and $j'_1=0$, $j'_2=2$.
Calculations are done with the BMKP PES (bold lines), triangles up are corresponding 
results from work \cite{schaefer90}.}
\label{fig:fig3}
\end{figure}

\begin{figure}
\begin{center}
\includegraphics*[scale=1.0,width=27pc,height=17pc]{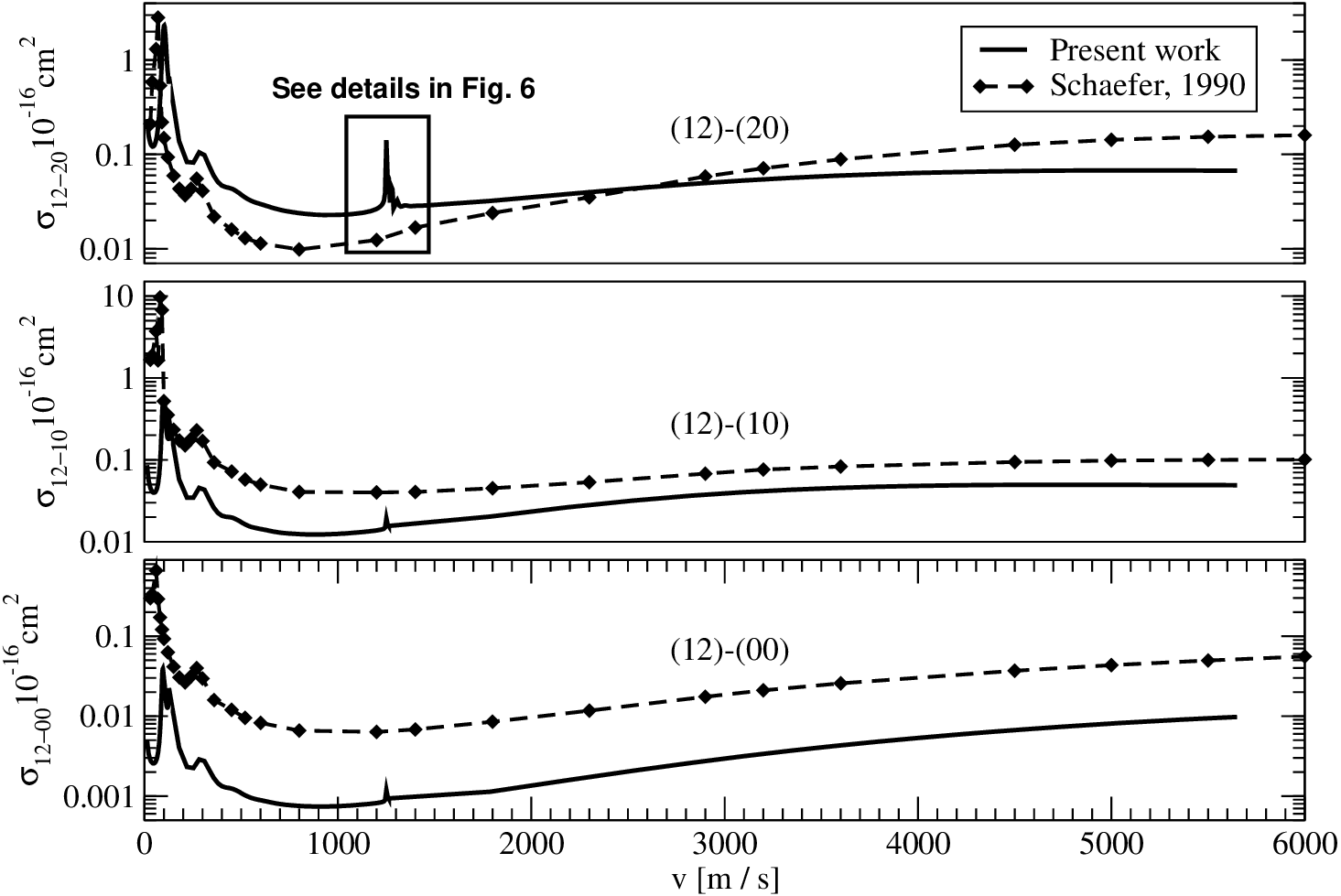}
\end{center}
\caption{Rotational state resolved integral cross sections for
$p$-$\mbox{H}_2(j_2) +\mbox{HD}(j_1) \rightarrow \mbox{H}_2(j'_2) + \mbox{HD}(j'_1)$.
The initial states of HD and H$_2$ molecules are $j_1=1$ and $j_2=2$ respectively and 
corresponding final states are $j'_1=2, j'_2=0$ in the upper, $j'_1=1$, $j'_2=0$ in the
midle, and $j'_1=j'_2=0$ in the bottom plots.
Calculations are done with the BMKP PES (bold lines), diamonds are corresponding 
results from work \cite{schaefer90}.}
\label{fig:fig4}
\end{figure}

\begin{figure}
\begin{center}
\includegraphics*[scale=1.0,width=27pc,height=17pc]{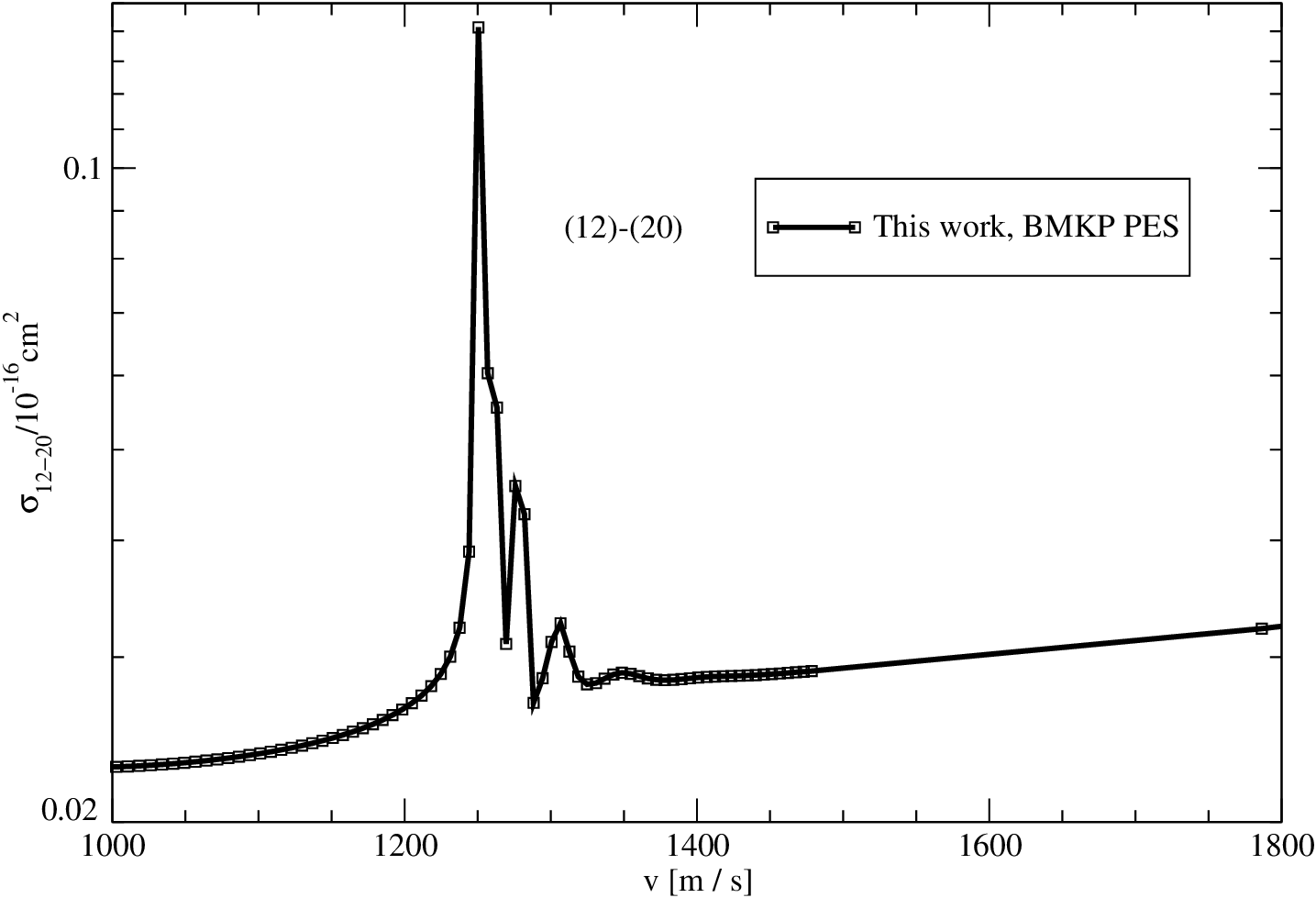}
\end{center}
\caption{Sharp resonance in the $(j_1=1,j_2=2)\rightarrow (j'_1=2,j'_2=0)$ rotational transition state cros section
of $p$-$\mbox{H}_2(j_2) +\mbox{HD}(j_1)$.    
The kinetic velocity ranges from 1000 m/s to 1800 m/s in this graph.}
\label{fig:fig5}
\end{figure}

\end{document}